\documentclass[aps,showpacs,preprintnumbers,amsmath,amssymb,twocolumn,superscriptaddress,prx,floatfix, nofootbib
]{revtex4-2}

\usepackage{amsmath}
\usepackage{graphicx}
\usepackage{dcolumn}
\usepackage{bm}
\usepackage{xr-hyper} 
\usepackage{hyperref}
\newcommand{\abs}[1]{\left\lvert #1 \right\rvert}
\usepackage[dvipsnames]{xcolor}

\begin{document}
\title{Description length of canonical and microcanonical models}

\author{Francesca Giuffrida}
\email{Contact Author: francesca.giuffrida@imtlucca.it}
\affiliation{IMT School for Advanced Studies, Lucca (Italy)}
\affiliation{Lorentz Institute for Theoretical Physics, Leiden University, Leiden (The Netherlands)}
\author{Tiziano Squartini}
\affiliation{IMT School for Advanced Studies, Lucca (Italy)}
\affiliation{INdAM-GNAMPA Istituto Nazionale di Alta Matematica (Italy)}
\affiliation{Scuola Normale Superiore, Pisa (Italy)}
\author{Peter Gr\"{u}nwald}
\affiliation{Centrum Wiskunde \& Informatica, Amsterdam (The Netherlands)}
\affiliation{Mathematical Institute, Leiden University, Leiden (The Netherlands)}
\author{Diego Garlaschelli}
\affiliation{IMT School for Advanced Studies, Lucca (Italy)}
\affiliation{Lorentz Institute for Theoretical Physics, Leiden University, Leiden (The Netherlands)}
\affiliation{INdAM-GNAMPA Istituto Nazionale di Alta Matematica (Italy)}

\begin{abstract}
The (non-)equivalence of canonical and microcanonical ensembles is a fundamental question in statistical physics, concerning whether the use of soft and hard constraints in the maximum-entropy construction leads to the same description of a system. Despite the fact that maximum-entropy models are also commonly used in statistical inference, pattern detection, and hypothesis testing, a complete understanding of the effects of ensemble non-equivalence on statistical modeling is still missing. Here, we study this problem from a rigorous model selection perspective by comparing canonical and microcanonical models via the Minimum Description Length (MDL) principle, which yields a trade-off between likelihood, measuring model accuracy, and complexity, measuring model flexibility and its potential to overfit data. We compute the Normalized Maximum Likelihood (NML) of both formulations and find that: (i) microcanonical models always achieve higher likelihood but are always more complex; (ii) the optimal model choice depends on the empirical values of the constraints -- the canonical model performs best when its fit to the observed data exceeds its uniform average fit across all realizations; (iii) in the thermodynamic limit, the difference in description length per node vanishes when ensemble equivalence holds but persists otherwise, showing that
non-equivalence implies extensive differences between large canonical and microcanonical models. Finally, we compare the NML approach to Bayesian methods, showing that (iv) the choice of priors, practically irrelevant in equivalent models, becomes crucial when an extensive number of constraints is enforced, possibly leading to very different outcomes.
\end{abstract}

\maketitle

\section{Introduction}

Entropy maximization provides a principled approach to inference, resulting in maximally random models under specified constraints. These models have found applications across a wide range of scientific disciplines, such as network science~\cite{squartini_springer,cimini19,squartini18} and time-series analysis~\cite{marcaccioli2020maximum,marcaccioli2020correspondence}.

Two distinct formulations of maximum-entropy models are possible depending on how constraints are enforced. Canonical models are defined by requiring the value of the constraints to be satisfied \emph{on average}. Microcanonical models are defined by requiring the value of the constraints to be satisfied \emph{exactly}. These two formulations do not necessarily lead to the same description of a given system, i.e., they are, in general, \emph{non-equivalent}.

 The concept of ensemble equivalence, introduced by Gibbs~\cite{gibbs}, plays an important role in statistical mechanics. Traditional studies have shown that canonical and microcanonical ensembles tend to become equivalent in the thermodynamic limit for systems with short-range interactions. Violations of this asymptotic equivalence, known as ensemble non-equivalence, have historically been linked to systems with long-range interactions or other forms of non-additivity, such as mean-field spin models or self-gravitating systems~\cite{ blume71, lynden99, ellis2000, dagostino2000, barre2001, chavanis03, ellis04, bouchet2005, barre07, radin13,  campa2014, touchette15}.

More recently, it has become clear that ensemble non-equivalence can also arise due to an additional mechanism: the presence of an \emph{extensive number of constraints}~\cite{squartini15, squartini17, qi, garlaschelli18}. This has been demonstrated, for example, in network models with local constraints, such as the Configuration Model. In these cases -- unlike the classical examples from statistical physics -- non-equivalence is not necessarily associated with long-range interactions or phase transitions, and is not confined to specific parameter regimes, but is present across the entire parameter space. Moreover, this phenomenon can have practical implications for the expected macroscopic properties of the system under study (see~\cite{bruno20} for an example concerning bipartite networks).

Whenever the two approaches are equivalent, they identify the same set of typical configurations. In these cases, preferring one formulation over the other becomes a matter of convenience (e.g., computational efficiency). Whenever the two approaches are non-equivalent, the choice depends on the specific characteristics of the system under study and the nature of the constraints involved. For example, in the context of choosing a null model to be compared with real data, for e.g., pattern detection, it has been argued that, when non-equivalence holds, the ensemble choice should be based on a theoretical guiding principle~\cite{garlaschelli17, qi, squartini15bis, garlaschelli18, garlaschelli16}. 
In this case, if the data are expected to be noisy, meaning that the measured values of the constraints generally do not match the true (unobservable) ones, one should prefer the canonical version of the null model because it allows for fluctuations in the constraint quantities, while the microcanonical one would paradoxically assign zero probability to the true configuration of the system. Conversely, if, by hypothesis, the measured values of the constraints are not affected by error or noise, the microcanonical model should be preferred.

When there is no prior expectation available about the presence or absence of noise in the empirical values of the constraints, the decision between hard and soft constraints should be based on posterior evidence, i.e., which of the two null models best describes the data. This immediately leads to a problem of model selection between canonical and microcanonical ensembles.\\

Information theory provides a powerful framework for model selection by offering criteria that capture the trade-off between a model's ``goodness of fit,'' quantified by its log-likelihood, and its ``complexity.'' In general, model complexity measures the flexibility of a model in fitting different datasets. In other words, it measures how easily the model can overfit the data. A common way to quantify model complexity is by counting the number $k$ of the model's free parameters, which, in the case of maximum-entropy models, corresponds to the number of constraints. This is the case of the two historically most popular model selection criteria: the \emph{Akaike Information Criterion} (AIC)~\cite{aic} and the \emph{Bayesian Information Criterion} (BIC)~\cite{bic}, according to which one should pick the model minimizing
\begin{equation}\label{aic}
\text{AIC} = -\ln \mathcal{L}+k
\end{equation}
and
\begin{equation}\label{bic}
\text{BIC}=-\ln\mathcal{L}+\frac{k}{2}\ln V,
\end{equation}
respectively, where $\mathcal{L}$ is the model likelihood and $V$ is the sample size. Canonical and microcanonical models defined by the same set of constraints share the same number of parameters. Therefore, according to both AIC and BIC, their complexity is the same, and a comparison based on these criteria would reduce to a mere comparison between their log-likelihood functions. 
This would naively lead to the conclusion that the microcanonical model is always the best-scoring one since it has the highest likelihood.
However, neither AIC nor BIC can be computed for microcanonical models since they are derived under regularity assumptions that are not even remotely met by the latter. Moreover, both criteria are asymptotic results derived under the assumption of a finite number of parameters as the system size approaches infinity~\cite{burnham}. This assumption is not met by models whose number of parameters scales with the size of the system.\\

To overcome these issues, here we consider the Minimum Description Length (MDL) principle~\cite{grunwald_book,grunwald2019, yamanishi2023}, a family of information-theoretic approaches to model selection seeking the model that provides the shortest description of data. Specifically, we focus on the approach based on the Normalized Maximum Likelihood (NML), or Shtarkov, distribution~\cite{rissanen96}, with the aim of investigating the trade-off between the difference of log-likelihoods and that of complexities in light of ensemble non-equivalence. 

 A particularly useful definition of non-equivalence, naturally connected to information theory, is the \emph{measure-level} one, which involves the Kullback-Leibler (KL) divergence between the microcanonical and canonical probability distributions. Under mild conditions, this approach is equivalent to other definitions commonly used in statistical physics~\cite{touchette15}. According to this definition, (non-)equivalence is characterized by a (non-)vanishing relative entropy density as the system size approaches infinity. Interestingly, the KL divergence between canonical and microcanonical distributions coincides with the difference between the corresponding log-likelihood functions~\cite{squartini15, squartini17}. Thus, measure-level (non-)equivalence can be inspected by simply considering the asymptotic behavior of this log-likelihood difference~\cite{squartini15}, providing a natural connection to statistical inference.

While the impact of non-equivalence on the difference between log-likelihoods is relatively well understood~\cite{squartini15, qi}, its effects on the difference between complexities remain largely unexplored. To encompass a broad spectrum of applications, our study focuses on maximum-entropy models suitable for studying systems represented by binary rectangular matrices (e.g., bipartite networks, time series). More precisely, we aim at determining i) whether the influence of non-equivalence persists even when the complexity term is accounted for, and ii) whether non-equivalent canonical and microcanonical models should, in fact, be regarded as statistically distinct models. We conclude that the answer to both questions is yes, raising the question of whether non-equivalence can be defined at the model selection level. Furthermore, we show that in the presence of extensively many constraints, the choice of prior in a Bayesian framework can critically affect the resulting description length, underscoring the importance of this choice in non-equivalent settings.

The rest of the paper is organized as follows. Section~\ref{mdl} introduces MDL for discrete data and the Normalized Maximum Likelihood (NML) approach to define description lengths. In Section~\ref{dl}, we provide a general expression for the difference between the description lengths of canonical and microcanonical models as a function of the Kullback-Leibler divergence between the two distributions. In Section~\ref{app}, we explicitly derive and compare the description lengths of both the canonical and the microcanonical formulations of models defined by one global constraint (i.e., the sum of the entries of a matrix) as well as one-sided local constraints (i.e., the row-specific sums of a matrix). While the first constraint leads to ensemble equivalence, the second set of constraints leads to ensemble non-equivalence. Finally, Section~\ref{nml_bayes} reviews the relationship between the NML-based and Bayesian approaches when non-equivalence holds.

\section{The Minimum Description Length principle}\label{mdl}

The MDL principle embodies the idea that the best model to describe a given dataset is the one providing the shortest description. We now introduce the basic formalism by focusing on the discrete data $\bm{x}\in\mathcal{X}$, where $\mathcal{X}$ represents the space of all possible samples of fixed size $V$. A parametric statistical model $\mathcal{M}$ consists of a family of probability distributions sharing the same functional form, i.e.,
\begin{equation}
\mathcal{M}=\{P_{\mathcal{M}}(\bm{x};\bm{\theta})\}_{\bm{\theta}\in\Theta}
\end{equation}
where $\bm{\theta}$ is the vector of parameters whose values lie in the set $\Theta$.

In order to apply the MDL principle, one needs to compute the description length of the data $\bm{x}$ provided by a model $\mathcal{M}$. By Kraft's inequality~\cite{grunwald_book, li2008}, the optimal number of natural digits (nats) needed to describe the outcome $\bm{x}$ of a probability distribution $P(\bm{x})$ is given by $-\ln P(\bm{x})$. Since, however, $\mathcal{M}$ encompasses many distributions corresponding to different parameter values, a natural choice is that of considering the description length provided by the distribution minimizing the quantity $-\ln P_{\mathcal{M}}(\bm{x};\bm{\theta})$, reading
\begin{equation}
\mathcal{L}_{\mathcal{M}}(\bm{x})\equiv P_{\mathcal{M}}(\bm{x};\hat{\bm{\theta}}(\bm{x})),
\end{equation}
with $\hat{\bm{\theta}}(\bm{x})$ being the maximum-likelihood (ML) estimator of $\bm{\theta}$, i.e.
\begin{equation}
\hat{\bm{\theta}}(\bm{x})=\arg\max_{\bm{\theta}\in\Theta}P_{\mathcal{M}}(\bm{x};\bm{\theta}).
\end{equation}

However, the distribution $\mathcal{L}_{\mathcal{M}}(\bm{x})$ is affected by (at least) two problems having the same origin: the parameters appearing in its definition are evaluated for each individual data in $\mathcal{X}$. As a consequence, i) $\mathcal{L}_{\mathcal{M}}(\bm{x})$ cannot be a probability distribution with support in $\mathcal{X}$ as it is not properly normalized, and ii) adopting $\mathcal{L}_{\mathcal{M}}(\bm{x})$ may lead to severe overfitting since it is defined \emph{a posteriori} relative to the observations.
 
To solve these problems, one can build a so-called \textit{universal distribution} relative to $\mathcal{M}$, i.e., a unique, representative distribution $\bar{P}_{\mathcal{M}}(\bm{x})$ of the statistical model $\mathcal{M}$. Once equipped with a universal distribution, the description length of $\bm{x}$ through is $\mathcal{M}$ is defined as
\begin{equation}
\text{DL}^{\bar{P}}_{\mathcal{M}}(\bm{x}) \equiv -\ln \bar{P}_{\mathcal{M}}(\bm{x}).
\end{equation}
Universal distributions (which don't necessarily belong to $\mathcal{M}$) can be defined in several ways. In what follows, we will consider the MDL recipe based upon the NML universal distribution (for a more formal introduction to universal coding, we redirect the interested reader to~\cite{grunwald_book}).

\subsection{The Normalized Maximum Likelihood universal distribution} 

The NML distribution corresponds to the distribution attaining the minimum point-wise \emph{regret} in the worst-case scenario~\cite{grunwald_book}, i.e.,
\begin{equation}\label{minimax}
\bar{P}^\textrm{NML}_{\mathcal{M}}\equiv\arg\min_{\bar{P}}\max_{\bm{x}\in\mathcal{X}} \left\{-\ln\bar{P}(\bm{x})-[-\ln P_{\mathcal{M}}(\bm{x}; \hat{\bm{\theta}}(\bm{x}))]\right\}.
\end{equation}
In words, it is the distribution that \emph{a priori} requires the minimum number of nats when compared to the shortest description length \emph{a posteriori}, that is $-\ln P_{\mathcal{M}}(\bm{x};\hat{\bm{\theta}}(\bm{x}))$. The minimum is found in the worst-case scenario, i.e., for the data $\bm{x}$ for which this difference is maximal. The unique solution to this \emph{minimax} problem reads
\begin{equation}
\bar{P}^\textrm{NML}_{\mathcal{M}}(\bm{x})=\frac{P_{\mathcal{M}}(\bm{x};\hat{\bm{\theta}}(\bm{x}))}{\sum_{\bm{y}\in \mathcal{X}}P_{\mathcal{M}}(\bm{y};\hat{\bm{\theta}}(\bm{y}))},
\end{equation}
which is a normalized maximum likelihood. Following the universal coding approach, the description length of $\bm{x}$ through model $\mathcal{M}$ is computed as
\begin{align}\label{nml1}
\text{DL}^\textrm{NML}_\mathcal{M}(\bm{x})\equiv&-\ln\bar{P}^\textrm{NML}_{\mathcal{M}} (\bm{x})\nonumber\\
=&-\ln P_{\mathcal{M}}(\bm{x};\hat{\bm{\theta}}(\bm{x}))+\ln\sum_{\bm{y}\in \mathcal{X}}P_{\mathcal{M}}(\bm{y};\hat{\bm{\theta}}(\bm{y})).
\end{align}
While the first term is (minus) a model maximum log-likelihood, the second term
\begin{equation}\label{comp1}
\text{COMP}_\mathcal{M}\equiv\ln\sum_{\bm{y}\in\mathcal{X}}P_\mathcal{M}(\bm{y};\hat{\bm{\theta}}(\bm{y}))=\ln\sum_{\bm{y}\in \mathcal{X}}\mathcal{L}_\mathcal{M}(\bm{y})
\end{equation}
is named \emph{parametric complexity} of $\mathcal{M}$. Notice that $\mathcal{X}$ represents the set of all possible data sets of fixed size $V$, i.e., all the observable samples of size $V$. Thus, $\text{COMP}_\mathcal{M}$ depends on both the model and $\mathcal{X}$. Finally, we can rewrite equation (\ref{nml1}) as
\begin{equation}\label{dl_nml}
\text{DL}^\textrm{NML}_\mathcal{M}(\bm{x})=-\ln\mathcal{L}_{\mathcal{M}}(\bm{x})+\text{COMP}_\mathcal{M}.
\end{equation}
Similarly to other criteria rooted in information theory, this description length consists of two terms: the log-likelihood term, evaluated on the data and favoring models with a high goodness-of-fit, and the complexity term, depending only on the chosen model and penalizing the ones that are too complex.

Once we have evaluated the description length of our data according to a basket of competing models, the MDL principle prescribes selecting the one providing the minimum DL.

\section{Description length of canonical and microcanonical models} \label{dl}

Although computing the complexity term within the NML-based description length can be challenging, models defined by a sufficient statistic admit a simpler expression for this term. This section introduces an important class of models of this kind, the maximum-entropy models (MEMs), both in their canonical and microcanonical formulations.

Specifically, we will focus on discrete data that can be represented as a binary $n\times m$ matrix $\mathbf{G}$, i.e.
\begin{equation}
\mathbf{G}=\{g_{ij}\}_{\substack{i=\{1\dots n\}\\j=\{1\dots m\}}}
\end{equation}
and denote by $\mathcal{G}$ the set of all such matrices that, when endowed with a probability distribution $P(\mathbf{G})$, constitute a proper ensemble of matrices. The matrix representation is often used to describe systems composed of $n$ elements to be modeled (corresponding to the number of rows), each characterized by $m$ state variables or degrees of freedom (corresponding to the number of columns). The size of the data set is fixed and supposed to represent the number of independent entries of the matrix, i.e., $n m$. 

In the most common scenario, we can access a vector of quantities $\mathbf{c}^*=\mathbf{c}(\mathbf{G}^*)$, evaluated on the only available matrix $\mathbf{G}^*$, and representing the sufficient statistic of the model. Maximum-entropy models are, then, obtained by looking for the probability distribution maximizing Shannon entropy
\begin{equation}\label{ent_func}
\mathcal{S}[P]=-\sum_{\mathbf{G}\in\mathcal{G}}P(\mathbf{G})\ln P(\mathbf{G})
\end{equation}
while constraining the sufficient statistics. If the model is required to reproduce the value of the constraints exactly, i.e., $\mathbf{c}(\mathbf{G})=\mathbf{c}^*$, one obtains a \emph{microcanonical} model. If, instead, the model is required to reproduce the value of the constraints on average, i.e., $\langle\mathbf{c}(\mathbf{G})\rangle_P=\mathbf{c}^*$ with $\langle\cdot \rangle_P$ being the ensemble average induced by distribution $P$, one obtains the corresponding \emph{canonical} formulation. This approach, introduced by Gibbs~\cite{gibbs} and reprised by Jaynes~\cite{jaynes}, leads to ensembles of matrices that reproduce (either exactly or on average) the constrained quantities while randomizing everything else.

In what follows, we will adopt the NML-based approach to compute the description lengths of microcanonical and canonical MEMs, which is optimal (in a minimax sense) when assuming no prior knowledge on the parameters. Thus, the description length is taken as the one defined in equation (\ref{dl_nml}). For models admitting a sufficient statistic, the complexity term takes the convenient form
\begin{equation}\label{mem_comp}
\text{COMP}=\ln\sum_{\mathbf{c}\in\mathcal{C}}\Omega({\mathbf{c}})\mathcal{L}(\mathbf{c}),
\end{equation}
where $\mathcal{L}(\mathbf{c})$ denotes the value of the maximum likelihood in correspondence with a sufficient statistic reading $\mathbf{c}$ and $\Omega(\mathbf{c})$ denotes the number of configurations with $\mathbf{c}(\mathbf{G})=\mathbf{c}$, i.e.,
\begin{equation} \label{omega_def}
\Omega(\mathbf{c})\equiv\sum_{\mathbf{G}\in\mathcal{G}}\delta_{\mathbf{c}(\mathbf{G}),\mathbf{c}}
\end{equation}
with $\delta_{i,j}$ being the Kronecker delta. As this set may be empty for some values of $\mathbf{c}$, we will solely focus on the \textit{graphical values}, i.e., the set $\mathcal{C}=\{\mathbf{c}:\Omega(\mathbf{c})\neq\emptyset\}$ of values that are verified by at least one configuration, and denote it with $\mathcal{N}_{\mathcal{C}}=|\mathcal{C}|$ its cardinality. In other words, the sum in equation (\ref{mem_comp}) runs over the graphical values of the sufficient statistics rather than over all possible data.

\subsection{Description length of microcanonical models}\label{sec_dl_mic}  

When considering microcanonical models, entropy maximization yields the following functional form
\begin{equation}\label{p_mic}
P_\textrm{mic}(\mathbf{G};\mathbf{c})=
\begin{cases}
\frac{1}{\Omega(\mathbf{c})} & \text{if}\:\mathbf{c}(\mathbf{G})=\mathbf{c}\\
0 & \text{else},
\end{cases}
\end{equation}
where $\mathbf{c}$ is the vector of parameters whose values lie in the discrete set $\mathcal{C}$ - for microcanonical models, in fact, the vector of parameters corresponds to the sufficient statistics itself - and $P_\textrm{mic}(\mathbf{G};\mathbf{c})$ is a uniform probability whose mass differs from zero only in correspondence of those configurations that verify the condition $\mathbf{c}(\mathbf{G})= \mathbf{c}$. The derivation of this probability functional form as the solution of entropy maximization under hard constraints is given in more detail in Section \ref{micro_as_MEM} of the Appendix.

We will now evaluate the description length of a microcanonical model. Its maximum log-likelihood simply reads
\begin{equation}
\ln\mathcal{L}_\textrm{mic}(\mathbf{c})=\ln\frac{1}{\Omega(\mathbf{c})} = -\ln \Omega(\mathbf{c}).
\end{equation}
While combining this equation with equation (\ref{mem_comp}), one obtains the following expression for the related complexity:
\begin{equation} \label{mic_comp}
\text{COMP}_\textrm{mic}=\ln\sum_{\mathbf{c}\in\mathcal{C}}\Omega(\mathbf{c})\cdot\frac{1}{\Omega(\mathbf{c})}=\ln\sum_{\mathbf{c}\in\mathcal{C}}1=\ln\mathcal{N}_\mathcal{C}.
\end{equation}
Hence, the description length of a microcanonical maximum-entropy model is
\begin{equation}\label{mic_nml_dl}
\text{DL}_\textrm{mic}^\text{NML}(\mathbf{G}^*)=\ln\Omega(\mathbf{c}^*)+\ln \mathcal{N}_\mathcal{C}.
\end{equation}

\subsection{Description length of canonical models}

When considering canonical models, entropy maximization yields the following well-known exponential form
\begin{equation}\label{p_can}
P_\textrm{can}(\mathbf{G};\bm{\theta})=\frac{e^{-\bm{\theta}\cdot\mathbf{c}(\mathbf{G})}}{Z(\bm{\theta})}
\end{equation}
with $Z(\bm{\theta})=\sum_{\mathbf{G}\in\mathcal{G}} e^{-\bm{\theta}\cdot \mathbf{c}(\mathbf{G})}$ being the \textit{partition function}. In the context of network theory, the maximum-entropy canonical models are also known as \emph{Exponential Random Graph Models}~\cite{holland1981,hunter2006, cimini19}. In this case, the vector of parameters $\bm{\theta}$ corresponds to the vector of Lagrange multipliers needed to carry out a constrained entropy maximization. The ML estimators $\bm{\hat{\theta}}(\mathbf{G})$ are found by solving the system of equations
\begin{equation}\label{canonical_constraints}
\langle\mathbf{c}\rangle_{\bm{\hat{\theta}}(\mathbf{G})}=\mathbf{c}(\mathbf{G})
\end{equation}
with $\langle\cdot\rangle_{\bm{\theta}}$ representing the ensemble average induced by $P_\textrm{can}(\mathbf{G};\bm{\theta})$.

The maximum log-likelihood of a canonical model is linked to the microcanonical one through the Kullback-Leibler divergence:
\begin{equation}
\left. S(P_\textrm{mic}||P_\textrm{can}) \right|_{\mathbf{c},\bm{\theta}}=\sum_{\mathbf{G}\in\mathcal{G}} P_\textrm{mic}(\mathbf{G};\mathbf{c}) \ln\frac{P_\textrm{mic}(\mathbf{G};\mathbf{c})}{P_\textrm{can}(\mathbf{G};\bm{\theta})}.
\end{equation}
\noindent In fact, it can be shown~\cite{squartini17,squartini15} that
\begin{align}\label{DKL}
D_\text{KL}(\mathbf{c^*})&\equiv S(P_\textrm{mic}||P_\textrm{can})|_{\mathbf{c}^*,\bm{\theta}^*}\nonumber\\ &=\ln \frac{P_\textrm{mic}(\mathbf{G}^*;\mathbf{c}^*)}{P_\textrm{can}(\mathbf{G}^*;\bm{\theta}^*)}\nonumber\\
&=\ln\mathcal{L}_\textrm{mic}(\mathbf{c}^*)-\ln\mathcal{L}_\textrm{can}(\mathbf{c}^*)
\end{align}
where $\bm{\theta}^*=\hat{\bm{\theta}}(\mathbf{c}^*)$. Hence,
\begin{align}\label{mic_can}
\mathcal{L}_\textrm{can}(\mathbf{c})&=\mathcal{L}_\textrm{mic}(\mathbf{c})\cdot e^{-D_\text{KL}(\mathbf{c})}=\frac{1}{\Omega(\mathbf{c})}\cdot e^{-D_\text{KL}(\mathbf{c})}
\end{align}
for any value of the sufficient statistic, and
\begin{align}\label{can_comp}
\text{COMP}_\textrm{can}^\text{NML}&=\ln \sum_{\mathbf{c}\in \mathcal{C}} \Omega(\mathbf{c})\cdot\frac{1}{\Omega(\mathbf{c})}\cdot e^{-D_\text{KL}(\mathbf{c})}\nonumber \\&=\ln\sum_{\mathbf{c}\in\mathcal{C}}e^{-D_\text{KL}(\mathbf{c})}.
\end{align}

Consequently, the description length of a canonical model can be expressed as
\begin{align}
\text{DL}_\textrm{can}^\text{NML}(\mathbf{G}^*)&=-\ln\mathcal{L}_\textrm{can}(\mathbf{c}^*)+\ln\sum_{\mathbf{c}\in \mathcal{C}}e^{-D_\text{KL}(\mathbf{c})}\nonumber\\
&=\ln \Omega(\mathbf{c}^*)+D_\text{KL}(\mathbf{c}^*)+\ln \sum_{\mathbf{c}\in \mathcal{C}}e^{-D_\text{KL}(\mathbf{c})}.
\end{align}

\subsection{Comparing canonical and microcanonical models}\label{comparing}

We can now compare the description length of canonical and microcanonical models. Given that $D_\text{KL}$ is guaranteed to be non-negative, we have
\begin{align}\label{delta_L}
\Delta\ln\mathcal{L}(\mathbf{G}^*)&\equiv\ln\mathcal{L}_\textrm{can}(\mathbf{G}^*)-\ln\mathcal{L}_\textrm{mic}(\mathbf{G}^*)\nonumber\\
&=-D_\text{KL}(\mathbf{c}^*)\leq 0
\end{align} 
i.e., the canonical likelihood is always smaller than or equal to the microcanonical one. For the same reason, it is straightforward to see that the canonical complexity is always smaller than, or equal to, the microcanonical one:
\begin{align}
\Delta\text{COMP}&\equiv\text{COMP}_\textrm{can}-\text{COMP}_\textrm{mic}\nonumber\\
&=\ln\frac{\sum_{\mathbf{c}\in\mathcal{C}}e^{-D_{KL}(\mathbf{c})}}{\mathcal{N}_\mathcal{C}}\leq 0.
\end{align}
Thus, microcanonical models achieve a higher goodness-of-fit but, at the same time, are more complex. The interplay between the two differences determines the difference between the canonical and the microcanonical description length:
\begin{align}\label{thedifference}
\Delta\text{DL}(\mathbf{G}^*)&\equiv\text{DL}^{\text{NML}}_\textrm{can}(\mathbf{G}^*)-\text{DL}^{\text{NML}}_\textrm{mic}(\mathbf{G}^*)\nonumber\\
&=-\Delta\ln\mathcal{L}(\mathbf{G}^*)+\Delta\text{COMP}\nonumber\\
&=D_\text{KL}(\mathbf{c}^*)+\ln\frac{\sum_{\mathbf{c}\in\mathcal{C}}e^{-D_\text{KL}(\mathbf{c})}}{\mathcal{N}_\mathcal{C}}.
\end{align}

The following reasoning provides a different way of looking at the expression above. The canonical distribution $P_{\text{can}}(\mathbf{G};\bm{\theta})$ induces the following distribution on the values of the sufficient statistic
\begin{equation}
Q_{\text{can}}(\mathbf{c};\bm{\theta})=\Omega(\mathbf{c})P_{\text{can}}(\mathbf{c};\bm{\theta})
\end{equation}
where $P_{\text{can}}(\mathbf{c};\bm{\theta})$ is the canonical probability of a matrix $\mathbf{G}$, such that $\mathbf{c}(\mathbf{G})=\mathbf{c}$, for a given value of the vector of parameters $\bm{\theta}$. The maximum likelihood assigned to $\mathbf{c}$ by the correspondent model $\{Q_{\text{can}}(\mathbf{c},\bm{\theta})\}_{\theta \in \Theta}$ is, thus,
\begin{align}
Q_{\text{can}}(\mathbf{c};\hat{\bm{\theta}}(\mathbf{c}))&=\Omega(\mathbf{c}) P_{\text{can}}(\mathbf{c}; \hat{\bm{\theta}}(\bm{c})) \nonumber \\&=\frac{\mathcal{L}_{\text{can}}(\mathbf{c})}{\mathcal{L}_{\text{mic}}(\mathbf{c})}=e^{-D_\text{KL}(\mathbf{c})}
\end{align}
and
\begin{multline}
\Delta\text{DL}(\mathbf{G}^*)=-\ln Q_{\text{can}}(\mathbf{c}^*;\bm{\theta}^*)\\+\ln \frac{\sum_{\mathbf{c}\in\mathcal{C}}Q_{\text{can}}(\mathbf{c};\hat{\bm{\theta}}(\mathbf{c}))}{\mathcal{N}_\mathcal{C}};
\end{multline}
upon defining
\begin{equation}
\bar{Q}_{\text{can}}\equiv \frac{\sum_{\mathbf{c}\in\mathcal{C}}Q_{\text{can}}(\mathbf{c},\hat{\bm{\theta}}(\mathbf{c}))}{\mathcal{N}_\mathcal{C}} 
\end{equation}
as the uniform average of $Q_{\text{can}}(\mathbf{c};\hat{\bm{\theta}}(\mathbf{c}))$ over all possible values of $\mathbf{c}\in \mathcal{C}$, we find that the difference between description lengths can be expressed as
\begin{equation} \label{Q_def}
\Delta\text{DL}(\mathbf{G}^*)=\ln \frac{\bar{Q}_{\text{can}}}{Q_{\text{can}}^*},
\end{equation}
where $ Q_{\text{can}}^*\equiv Q_{\text{can}}(\mathbf{c}^*, \bm{\theta}^*)$. This expression depends only on the canonical model, providing a new interpretation of the comparison between the canonical and the microcanonical description length:
the canonical model provides a shorter description length whenever $\Delta\text{DL}(\mathbf{G}^*)<0$, i.e., whenever $Q_{\text{can}}^*>\bar{Q}_{\text{can}}$. In other words, the canonical model is the best-scoring model whenever the maximum likelihood it assigns to the observed value $\mathbf{c}^*$ is higher than the maximum likelihood averaged over all possible values of the sufficient statistic (or, more intuitively, if it performs `better than on average' on the observed sufficient statistic). Remarkably, this is in open contrast with a comparison based only on the likelihoods, according to which the microcanonical formulation should always be preferred, independently of the observed data.\\

Expression (\ref{thedifference}) shows that the entity of the difference between the two description lengths crucially depends on the value of the KL divergence. As said in the introduction, the canonical and the microcanonical ensembles are (non-)equivalent if the relative $\text{D}_\text{KL}$ is (non-)vanishing as the system size approaches infinity. When studying matrices, identifying the system size with the number of entries, i.e., $n m$, may be inappropriate as adding (only) one entry to a matrix without altering its structure is, in fact, impossible. In these cases, it may be much more reasonable to identify the system size with the number $n$ of items: increasing the system size from $n$ to $n+1$ would, thus, correspond to adding a node to a network or a time-point to a time series.

In this scenario, the canonical and the microcanonical ensembles are equivalent in the measure sense~\cite{touchette15} if

\begin{equation}\label{equiv_def}
\lim_{n\to\infty}\frac{D_\text{KL}(\mathbf{c}^*)}{n}= \lim_{n\to\infty}\frac{\abs{\Delta\ln\mathcal{L}(\mathbf{c}^*)}}{n} = 0
\end{equation} 
or, equivalently, if
\begin{equation}\label{equiv_delta_log}
D_{\text{KL}}(\mathbf{c}^*) = \abs{\Delta\ln\mathcal{L}(\mathbf{c}^*)}=o(n).
\end{equation}
where $o(x)$ stands for a quantity that goes to $0$ when divided by $x$ as $n\to + \infty$.

This definition of non-equivalence relies on comparing the likelihood functions of the two models. We aim to extend it in order to take into account the models' complexities. As a first step, we consider the order of $\Delta \text{DL}$. 
Notice that 
\begin{multline} \label{dkl_deltadl}
    D_{\text{KL}}(\mathbf{c}) = o(n) \quad \forall  \mathbf{c}\in \mathcal{C}\\  \quad \Rightarrow \quad \Delta \text{DL} (\mathbf{c}) = o(n) \quad \forall  \mathbf{c}\in \mathcal{C}
\end{multline} as it follows directly from (\ref{thedifference}). Thus, if the two ensembles are equivalent, the difference in description lengths between the corresponding models will be $o(n)$. However, a similar conclusion cannot be immediately derived when non-equivalence holds and $D_{\text{KL}}$ grows at least linearly with $n$. Indeed, the likelihood and complexity terms in (\ref{thedifference}) are of the same order and have opposite signs; therefore, evaluating the order of $\Delta \text{DL}$ is a non-trivial task in this case. The following section addresses this problem by explicitly evaluating $\Delta \text{DL}$ in two notable examples of maximum-entropy models.

\section{Applications to maximum-entropy models}\label{app}

The simplest constraint one may imagine to enforce is the sum of the elements of a matrix, i.e.
\begin{equation}\label{global_con}
l(\mathbf{G})\equiv\sum_{i=1}^n\sum_{j=1}^mg_{ij}.
\end{equation}
In the case of bipartite binary networks, this quantity corresponds to the total number of links and induces the well-known Erd\"os-R\'enyi model. While this model is one of the most popular ones, it is also widely recognized to fall short in capturing the topological heterogeneity that characterizes most real-world systems. An alternative approach to introduce heterogeneity is to constrain the sum of elements of each row, i.e.
\begin{equation}\label{local_con}
r_i(\mathbf{G})\equiv\sum_{j=1}^{m}g_{ij}\quad i=1\dots n.
\end{equation}
In the case of bipartite binary networks, the sequence $r_i(\mathbf{G})$, $i=1\dots n$ coincides with the degree sequence of the nodes belonging to just one layer and induces the canonical MEM known as Bipartite Partial Configuration Model~\cite{saracco17}.
As we will employ asymptotic results to compare the canonical and the microcanonical description lengths, we need to specify the scaling of the constraints $\mathbf{c}^*$ as $n \to \infty$. In what follows, we will focus on \emph{dense matrices}, i.e., we will assume that $g_{ij}^*=\Theta(1)$ $\forall (i, j)$, where $\Theta(x)$ indicates a quantity of the same order of $x$ as $n \to \infty$. This choice leads to $r_i^*=\Theta(m)$ and $l^*=\Theta(nm)$, the `actual' order of these quantities depending on the growth rate of $m$. We will consider two cases: (i) $m$ finite in the limit, i.e., $m = \Theta(1)$ and (ii) $m$ growing linearly with $n$, i.e., $m = \Theta(n)$. As a consequence of our assumption, the matrix density
\begin{align}
p^*=\frac{l^*}{nm}
\end{align}
and the row densities
\begin{align}
p_i^*=\frac{r_i^*}{m}\quad i=1\dots n
\end{align}
will remain finite in the asymptotic limit, i.e., $p^*=\Theta(1)$ and $p_i^*=\Theta(1)$, $\forall\:i$.

\subsection{Enforcing one global constraint}\label{global}

Since we are considering binary matrices, constraining the total number of $1$s in a microcanonical fashion leads to a number of configurations equal to $\Omega(l)=\binom{nm}{l}$. The corresponding microcanonical distribution, thus, becomes
\begin{equation}
P_{\text{mic}}(\mathbf{G};l)=
\begin{cases}
\frac{1}{\binom{nm}{l}} & \text{if}\:l(\mathbf{G})=l\\
0 & \text{else},
\end{cases}
\end{equation}
inducing the maximum log-likelihood
\begin{equation}
\ln\mathcal{L}_\textrm{mic}(l)=-\ln \binom{nm }{l}.
\end{equation}
The complexity equals the logarithm of the number of graphical values, i.e.
\begin{equation}
\text{COMP}_{\text{mic}}^\text{NML}=\sum_{l=0}^{nm}1=\ln[nm+1].
\end{equation}
The microcanonical description length, thus, reads
\begin{equation}\label{dl_mic_er}
\text{DL}_{\text{mic}}^\text{NML}(\mathbf{G}^*)=\ln\binom{nm}{l^*}+\ln[nm+1].
\end{equation}

The canonical probability can be computed explicitly, as this model represents one of the few canonical MEMs whose partition function can be computed analytically. Specifically, it reads $Z(\theta)=(1+e^{\theta})^{nm}$ and induces the expression
\begin{equation}
P_{\text{can}}(\mathbf{G};\theta)=\frac{e^{-\theta l(\mathbf{G})}}{(1+e^{-\theta})^{nm}}.
\end{equation}
Upon defining
\begin{equation}
p\equiv\frac{e^{-\theta}}{1+e^{-\theta}},
\end{equation}
we can re-write the canonical probability via the so-called \textit{mean-value parametrization} leading to the expression
\begin{equation}\label{bernoulli}
P_{\text{can}}(\mathbf{G};p)=p^{l(\mathbf{G})}(1-p)^{nm-l(\mathbf{G})}
\end{equation}
i.e., the distribution of a collection of $nm$ i.i.d. Bernoulli variables, indicating that each matrix element is either $1$, with probability $p$, or $0$, with probability $1-p$. The ML estimator of $p$ is given by the matrix density $\hat{p}(\mathbf{G})=l(\mathbf{G})/nm$ and the maximum log-likelihood, obtained by inserting $\hat{p}$ into (\ref{bernoulli}), reads
\begin{equation}
\ln\mathcal{L}_\textrm{can}(l)=-nm\cdot h\left(\frac{l}{nm}\right)
\end{equation}
where $h(p)=-p\ln(p)-(1-p)\ln(1-p)$ is the entropy of a Bernoulli distribution with a parameter equal to $p$. An exact expression for the canonical complexity has been derived in~\cite{staniczenko2014} and reads
\begin{equation}\label{comp_er}
\text{COMP}_\textrm{can}^\text{NML}=\ln\left[\frac{e^{nm}\Gamma(nm, nm)}{nm^{nm-1}}+1\right]
\end{equation}
where $\Gamma(s,x)$ is the upper incomplete gamma function. In case $s$ is a positive integer (as it happens to be our case), it can be expressed as
\begin{equation}
\Gamma(s,x)=e^{-x}(s-1)!\sum_{k=0}^{s-1}\frac{x^k}{k!}.
\end{equation}
In conclusion, the canonical description length reads
\begin{multline}\label{dl_can_er}
\text{DL}_{\text{can}}^\text{NML}(\mathbf{G}^*)=nm\cdot h\left(\frac{l^*}{nm}\right)\\+\ln\left[\frac{e^{nm}\Gamma(nm, nm)}{nm^{nm-1}}+1\right].
\end{multline}
Notably, the details of the matrix structure play no role as $\text{DL}_{\text{can}}^\text{NML}$ solely depend on global quantities, namely the number of $1$s and the sample size - in our case, $nm$. 

So far, all the reported results are exact. Nevertheless, as we are interested in the order of $\Delta \text{DL}$ when $n\to \infty$, we consider the following asymptotic results, derived from the exact ones (see section \ref{appendix_er} of the Appendix):
\begin{align}
\label{as_delta_l_er}\Delta\ln\mathcal{L}(\mathbf{G^*})&=-\frac{1}{2}\ln[nm]-\frac{1}{2}\ln[2\pi p^*(1-p^*)]+o(1),\\
\label{as_delta_c_er}\Delta\text{COMP}&=-\frac{1}{2}\ln[nm]+\frac{1}{2}\ln\left[\frac{\pi}{2}\right]+o(1)
\end{align}
and, upon combining them, the (asymptotic) difference between description lengths becomes
\begin{equation}
\label{as_delta_dl_er}\Delta \text{DL}(\mathbf{G}^*)=\frac{1}{2}\ln[\pi^2p^*(1-p^*)]+o(1).
\end{equation}

The expressions above provide two significant insights. First, the order of the difference between log-likelihood functions is $o(n)$, regardless of the growth rate of $m$, a result implying that condition (\ref{equiv_delta_log}) is met and ensemble equivalence holds. Second, the leading terms of expressions (\ref{as_delta_l_er}) and (\ref{as_delta_c_er}) are identical, thus canceling out when compared. Consequently, the (asymptotic) difference between description lengths depends on the size of the system only through the matrix density $p^*$, assumed to be finite in the dense regime. Thus, such a difference is finite in the same regime, in agreement with  (\ref{dkl_deltadl}).
\subsection{Enforcing \texorpdfstring{$n$}{n} local constraints}\label{local}

Let us now consider canonical and microcanonical models induced by an extensive number of parameters, each one corresponding to the sum of the elements of a different row. With this choice, the matrix can be viewed as a collection of $n$ independent strips of size $1\times m$. Since we impose a single constraint per row, all the prior results hold row-wise, with $m$ replacing $nm$ and $r_i$ replacing $l$. Specifically, the microcanonical probability distribution reads
\begin{equation}
P_{\text{mic}}(\mathbf{G};\mathbf{r})= 
\begin{cases}
\prod_{i=1}^n\frac{1}{\binom{m}{r_i}} &\text{if}\:\mathbf{r}(\mathbf{G})=\mathbf{r}\\
0 & \text{else,}
\end{cases}
\end{equation}
and the related maximum log-likelihood is the sum of the individual ones, i.e.,
\begin{equation}
\ln\mathcal{L}_\textrm{mic}(\mathbf{G}^*)=-\sum_{i=1}^n\ln\binom{m }{r_i^*}.
\end{equation}
Analogously, the microcanonical complexity reads
\begin{equation}
\text{COMP}_\textrm{mic}=n\ln[m+1],
\end{equation}
inducing the expression
\begin{equation}
\text{DL}_{\text{mic}}^\text{NML}(\mathbf{G}^*)=\sum_{i=1}^n \ln \binom{m}{r_i^*}+n\ln[m+1].
\end{equation}

Similarly, the canonical probability is the product of row-specific canonical probabilities
\begin{equation}\label{pcan_can}
P_{\text{can}}(\mathbf{G};\bm{\theta})=\prod_{i=1}^{n}\frac{e^{-\theta_ir_i(\mathbf{G})}}{(1+e^{-\theta_i})^m},
\end{equation}
and can be rewritten as 
\begin{equation}\label{bernoulli_pcm}
P_{\text{can}}(\mathbf{G};\mathbf{p})=\prod_{i=1}^n p_i^{r_i(\mathbf{G})}(1-p_i)^{m-r_i(\mathbf{G})}
\end{equation}
upon considering the mean-value parametrization
\begin{equation}
p_i \equiv \frac{e^{-\theta_i}}{(1+e^{-\theta_i})^m}\quad i=1\dots n.
\end{equation}
Since for each row the ML estimator of the parameter $p_i$ is given by the row density $\hat{p}_i(\mathbf{G})=\frac{r_i(\mathbf{G})}{m}$, the maximum log-likelihood reads

\begin{equation}
\ln\mathcal{L}_\textrm{can}(\mathbf{r})=-m\sum_{i=1}^{n}h\left(\frac{r_i}{m} \right)
\end{equation}
and the complexity reads
\begin{equation}\label{comp_pcm}
\text{COMP}_\textrm{can}=n\ln\left[\frac{e^{m}\Gamma(m, m)}{m^{m-1}}+1\right].
\end{equation}
In conclusion, the canonical description length reads
\begin{multline}\label{dl_can_pcm}
\text{DL}_{\text{can}}(\mathbf{G}^*)=m\sum_{i=1}^nh\left(\frac{r_i}{m}\right)\\+n\ln\left[\frac{e^{m}\Gamma(m, m)}{m^{m-1}}+1\right].
\end{multline}

In the hypothesis of $m$ growing linearly with $n$, i.e., $m=\Theta(n)$, we consider the following asymptotic expressions (for a derivation, see section \ref{appendix_pcm} of the Appendix)
\begin{align}
\label{delta_logl_pcm}
\Delta \ln\mathcal{L}(\mathbf{G^*})=&-\frac{n}{2}\ln[m]-\frac{1}{2}\sum_{i=1}^n\ln[2\pi  p_i^*(1-p_i^*)]\nonumber\\
&+\frac{n}{12m}-\frac{1}{12m}\sum_{i=1}^n\frac{1}{p_i^*(1-p_i^*)}+o(1),\\
\label{delta_comp_pcm}
\Delta \text{COMP}=&-\frac{n}{2}\ln[m]+\frac{n}{2}\ln\left[\frac{\pi}{2}\right]\nonumber\\
&+a\cdot\frac{n}{\sqrt{m}}+b\cdot\frac{n}{m}+o(1)
\end{align}
with
\begin{align*}
a&=\frac{2}{3}\sqrt{\frac{2}{\pi}}\simeq0.53,\\
b&=-\frac{11}{12}-\frac{4}{9\pi}\simeq-1.06.
\end{align*}
By combining these expressions, we obtain the asymptotic expression
\begin{align}\label{delta_dl_PCM}
\Delta\text{DL}^\text{}(\mathbf{G}^*)=&\frac{1}{2}\sum_{i=1}^n\ln[\pi^2 p_{i}^*(1-p_{i}^*)]+a\cdot\frac{n}{\sqrt{m}}\nonumber\\
&+c\cdot\frac{n}{m}+\frac{1}{12m}\sum_{i=1}^n\frac{1}{p_i^*(1-p_i^*)}+o(1)
\end{align}
where $c=-1-\frac{4}{9\pi}\simeq-1.14$.\\

As for the case of one global constraint, we focus on two aspects of these results. First, the difference between log-likelihood functions is either $\Theta(n)$ or $\Theta(n\ln n)$ depending on the order of $m$, i.e., $m=\Theta(1)$ in the first case and $m=\Theta(n)$ in the second case. In any case, according to (\ref{equiv_delta_log}), the ensemble equivalence is broken. In particular, if $m=\Theta(n)$, the leading terms of expressions (\ref{delta_logl_pcm}) and (\ref{delta_comp_pcm}) are both of order $\Theta(n\ln n)$, thus canceling out when compared, precisely as in the equivalent case. Nonetheless, the remaining terms are still of order $\Theta(n)$. Hence, the (asymptotic) difference between description lengths increases linearly with the system size, regardless of the growth rate of $m$.

\subsection{Future research: Towards model-level equivalence}

At this point, it is important to emphasize that comparing canonical and microcanonical \textit{ensembles} is not the same as comparing the corresponding \textit{models}. The former involves comparing two probability distributions, each obtained for specific parameter values, whereas the latter involves comparing two sets of distributions (i.e., two models), with each set sharing a common functional form within itself. While in the context of the measure-level definition of ensemble equivalence, the comparison between ensembles can be reduced to comparing their likelihood terms, the comparison between models must take into account both their likelihoods and complexities.

In this section, we have shown two examples where the leading-order terms of the differences in likelihood and complexity are identical. This confirms that the complexities play as crucial a role in model comparison as the likelihoods. Furthermore, we validated relation (\ref{dkl_deltadl}) in the case of a single global constraint: when ensemble equivalence holds, the difference in description lengths between the canonical and microcanonical ensembles turns out to be $o(n)$. At the same time, our result that $\Delta \text{DL}$ grows with $n$ in the case of $n$ local constraints shows that the signature of broken ensemble equivalence remains visible when model complexities are included in the comparison. 

These findings lead to the following question: can we generalize the definition of ensemble equivalence at the model level? To do so, we need a quantity that plays a role analogous to that of the relative entropy when comparing ensembles but which applies to the comparison of models.

$\Delta \text{DL}$ might seem like a good candidate but is not suitable for this purpose: while it does take into account the model complexities, it still depends on the observed values. To be more general, we propose a definition involving the universal NML distributions representing the two models. Inspired by the measure level definition of non-equivalence~\cite{touchette15} of equation \eqref{equiv_def},
we consider the following KL divergence
\begin{align}
D^{\text{NML}}_{\text{KL}} &\equiv S(P_{\text{mic}}^{\text{NML}}|| P_{\text{can}}^{\text{NML}}) \nonumber \\
&\nonumber= \sum_{\mathbf{G}\in\mathcal{G}} \bar{P}^{\text{NML}}_\textrm{mic}(\mathbf{G}) \ln\frac{\bar{P}^{\text{NML}}_\textrm{mic}(\mathbf{G})}{\bar{P}^{\text{NML}}_\textrm{can}(\mathbf{G})}\\
&\nonumber =  \sum_{\mathbf{G}\in\mathcal{G}} \bar{P}^{\text{NML}}_\textrm{mic}(\mathbf{G}) \Delta \text{DL}(\mathbf{G})\\ 
&\nonumber = \sum_{\mathbf{c}\in\mathcal{C}} \Omega(\mathbf{c}) \frac{1}{\Omega(\mathbf{c})\mathcal{N}_{\mathcal{C}}}\Delta \text{DL}(\mathbf{c}) \\
& =  \overline{\Delta \text{DL}(\mathbf{c})},
\end{align}    
where $\overline{\Delta \text{DL}(\mathbf{c})}$ represents the uniform average of $\Delta \text{DL}(\mathbf{c})$ over all possible values of $\mathbf{c} \in \mathcal{C}$. Following the standard definition and assuming that the more relevant scale to consider is still the system size $n$, we say that the canonical and microcanonical models are equivalent \textit{on the model level} if 
\begin{equation}\label{model_equiv_def}
\lim_{n\to\infty}\frac{D^{\text{NML}}_\text{KL}}{n}= \lim_{n\to\infty}\frac{\overline{\Delta \text{DL}(\mathbf{c})}}{n} = 0.
\end{equation}
Notice that, according to (\ref{dkl_deltadl}) 
\begin{equation} \label{dkl_dklnml}
    D_{\text{KL}}(\mathbf{c}) = o(n) \quad \forall  \mathbf{c}\in \mathcal{C} \quad \Rightarrow \quad  D_{\text{KL}}^{\text{NML}} = o(n).
\end{equation}
i.e., whenever ensemble equivalence holds, the corresponding microcanonical and canonical models are equivalent on the model level. The reverse implication, verified here in the specific case of one-sided local constraints, will be investigated in future work.

\section{NML-based VS Bayesian approach to model selection} \label{nml_bayes}

In the previous sections, we implicitly assumed that no prior knowledge about the parameters was available. Under this assumption, the NML-based universal distribution is the best choice in a precise minimax sense, as shown in section \ref{mdl}. Here, we consider the Bayesian approach to model selection~\cite{grunwald_book, grunwald2019} by introducing the Bayesian universal distribution
\begin{equation}\label{bayes_dist}
\bar{P}^\text{Bayes}_{\mathcal{M}}(\bm{x})=\int_{\Theta}P_{\mathcal{M}}(\bm{x}; \bm{\theta})w(\bm{\theta})d\bm{\theta}
\end{equation}
(see ~\cite{peixoto17, vallescatala18} for applications in the context of network models). The main difference with respect to the NML description length lies in the presence of a prior on the parameters $w(\bm{\theta})$ to be supplied by the user. This implementation, which is formally equivalent to the \textit{Bayes factor method}~\cite{kass95}, provides a description length for $\bm{x}$ through model $\mathcal{M}$ that is defined as
\begin{equation}\label{bayes_dl}
\text{DL}_\mathcal{M}^\text{Bayes}(\bm{x})\equiv-\ln \bar{P}^\text{Bayes}_{\mathcal{M}}(\bm{x}).
\end{equation}

In some cases, the NML-based approach can be retrieved in a Bayesian context once the Bayesian distribution is equipped with a prior such that the two description lengths are, at least asymptotically, the same. We will refer to these priors as \textit{NML-optimal priors}. In the microcanonical case, the uniform prior is an NML-optimal prior. Indeed, when the parameter space is a discrete set, the integral in (\ref{bayes_dl}) is replaced by a sum, and the microcanonical Bayesian description length reads
\begin{align}
\text{DL}_{\text{mic}}^\text{Bayes}(\mathbf{G}^*)&=-\ln \mathcal{L}_{\text{mic}}(\mathbf{G}^*)-\ln w(\mathbf{c}^*)\nonumber\\
&=\ln\Omega(\mathbf{c}^*)-\ln w(\mathbf{c}^*),
\end{align}
which coincides with the NML-based description length provided by equation (\ref{mic_nml_dl}), upon identifying $w(\mathbf{c})$ with a uniform prior over $\mathcal{C}$:
\begin{equation}
w^{\text{Uniform}}(\mathbf{c})=\frac{1}{\mathcal{N}_{\mathcal{C}}}\quad\forall\:\mathbf{c}\in\mathcal{C}.
\end{equation}

In the canonical case, the situation is quite different. First, consider $V$ i.i.d. outcomes $\bm{x} = (x_1, \dots x_V)$ from a general $k$-dimensional exponential model. When $\bm{x}$ is such that the ML estimators of $\bm{\theta}$ are bounded away from the boundaries of $\Theta$ as the sample size goes to infinity, and under some regularity conditions on $\mathcal{M}$ which hold in our setting, the following asymptotic results hold true:
\begin{multline}\label{as_formula_nml}
\text{DL}^\textrm{NML}_\mathcal{M}(\bm{x})=-\ln\mathcal{L}_{\mathcal{M}}(\bm{x})+\frac{k}{2}\ln\frac{V}{2\pi}\\
+\ln\int_{\Theta}\sqrt{\text{det}\mathbf{I}(\bm{\theta})}d\bm{\theta}+o(1),
\end{multline}
\begin{multline}\label{asymp_bayes}
\text{DL}_\mathcal{M}^\text{Bayes}(\bm{x})= -\ln\mathcal{L}_{\mathcal{M}}(\bm{x})+\frac{k}{2}\ln\frac{V}{2\pi}\\
-\ln w(\hat{\bm{\theta}}(\bm{x}))+\ln \sqrt{\text{det}\mathbf{I}(\hat{\bm{\theta}}(\bm{x}))}+o(1),
\end{multline}
where $\text{det}\mathbf{I}(\theta)$ is the determinant of the $k \times k$ \emph{normalized Fisher information matrix}. 

If we equip the Bayesian universal distribution with the Jeffreys prior~\cite{jeffreys1946}
\begin{align}\label{jeffreys}
J(\bm{\theta})=\frac{\sqrt{\text{det}\mathbf{I}(\bm{\theta})}}{\int_{\Theta} \sqrt{\text{det}\mathbf{I}(\bm{\theta})} d\bm{\theta}}
\end{align}
by posing $w(\bm{\theta})\equiv J(\bm{\theta})$, the two asymptotic expressions coincide (for a formal derivation of these results, we redirect the interested reader to ~\cite{rissanen96} and ~\cite{grunwald_book}). Thus, whenever the asymptotic formulas (\ref{as_formula_nml}) and (\ref{asymp_bayes}) hold true, the NML-based description length is asymptotically equivalent to the one provided by the Bayesian distribution equipped with Jeffreys prior (hereby, Bayes-Jeffreys description), which is NML-optimal.

This is verified in the case of one global constraint, where the resulting canonical model turns out to be an exponential i.i.d. model. The asymptotic formulas do not provide the rate of convergence between the two description lengths, which can be derived by directly comparing the exact expressions of the NML description length $\text{DL}_{\text{can}}^{\text{NML}}$ (\ref{dl_can_pcm}) and the Bayes-Jeffreys description length
\begin{multline}
\label{dl_jeff_er}
\text{DL}_{\text{can}}^{\text{Bayes-Jeffreys}}(\mathbf{G}^*) = \ln[\pi(nm)!]\\-\ln\left[\Gamma(nm-l^*+1/2)\Gamma(l^*+1/2)\right]
\end{multline}
(see section \ref{appendix_jeff} of the Appendix for a derivation). In the limit $n\to \infty$ one finds
\begin{multline}\label{as_diff_nml_jeff}
\text{DL}_{\text{can}}^{\text{NML}}(\mathbf{G}^*)-\text{DL}_{\text{can}}^{\text{Bayes-Jeffreys}}(\mathbf{G}^*)=\\ =\frac{2}{3}\sqrt{\frac{2}{\pi nm}} +\Theta\left(\frac{1}{nm}\right).
\end{multline}
The same comparison can be carried out in the case of $n$ local constraints by applying the results above to each row. In this case, the Bayes-Jeffreys description length reads
\begin{multline}
\text{DL}_{\text{can}}^{\text{Bayes-Jeffreys}}(\mathbf{G}^*)= n\ln[\pi m!] \\ - \sum_{i=1}^{n}\ln\left[\Gamma(m-r_i^*+1/2)\Gamma(r_i^*+1/2)\right]
\end{multline}
and in the limit $n\to \infty$ one finds
\begin{multline} \label{jeff_diff}
\text{DL}_{\text{can}}^{\text{NML}}(\mathbf{G}^*)-\text{DL}_{\text{can}}^{\text{Bayes-Jeffreys}}(\mathbf{G}^*)= \\= \frac{2}{3}\frac{\sqrt{2}n}{\sqrt{\pi m}}+\Theta \left(\frac{n}{m}\right).
\end{multline} 
As $n\to\infty$, the last difference diverges both in case $m$ stays finite, i.e., $m=\Theta(1)$, and in case $m$ diverges, upon assuming that the rate at which $m$ diverges does not exceed that of $n$. Consequently, in this case, it is no longer true that the NML approach can be asymptotically retrieved from the Bayesian one equipped with Jeffreys prior. Although the existence of a modified Jeffreys prior extending the equivalence of the two approaches to the case of $n$ local constraints can indeed be hypothesized, for the time being, we merely suggest not to assume this to be true when ensemble non-equivalence holds.\\

In the remainder of this section, we show that the Bayesian description length becomes more sensitive to the choice of prior whenever local constraints are enforced. Let us start by considering the relationship
\begin{equation}\label{nml_can_bayesian_mic}
\text{DL}_{\text{can}}^\text{NML} = \text{DL}_{\text{mic}}^\text{Bayes-Rissanen}
\end{equation}
stating that the NML-based description length of a canonical model coincides with the Bayesian description length of its microcanonical variant, equipped with the so-called \emph{canonical prior}, introduced by Rissanen in~\cite{rissanen2001} (see the definition in section \ref{ap_identities} of the Appendix). Moreover,
\begin{equation}\label{nml_mic_bayesian_can}
\text{DL}_{\text{mic}}^\text{NML}=\text{DL}_{\text{can}}^\text{Bayes-Uniform}
\end{equation}
i.e., the NML-based description length of the microcanonical model coincides with the Bayesian description length of its canonical variant equipped with a uniform prior on the mean-value parameters. 

The two identities, derived in \ref{ap_identities}, show that there exists a prior guaranteeing that the Bayesian description length of the canonical (microcanonical) model coincides with the NML-based description length of the microcanonical (canonical) model. This might lead to the conclusion that, in the framework of model selection, canonical and microcanonical models coincide even when non-equivalence holds if the right priors are chosen. Nevertheless, these priors are not NML-optimal. While this non-optimality is negligible in the case of one global constraint, it could play a crucial role when local constraints are enforced. In other words, in this case, different choices of prior can result in very different Bayesian description lengths for microcanonical as for canonical models. 

The $n$ local constraints case examined in the previous sections provides an instructive example. In this case, we showed that the difference between the canonical and microcanonical NML description lengths is $\Theta(n)$. Combining this result with (\ref{nml_can_bayesian_mic}), we can conclude that when it comes to the microcanonical model, choosing between the uniform prior (which is NML-optimal) and the canonical prior can amount to a difference in description length of $\Theta(n)$ nats.  Similarly, when it comes to the canonical model, it can be easily shown that, based on (\ref{jeff_diff}, \ref{nml_mic_bayesian_can}) when $m=\Theta(n)$, choosing between the uniform and the Jeffreys prior can lead to a difference in description length of $\Theta(\sqrt{n})$ nats. In contrast, the same differences are $\Theta(1)$ when one global constraint is involved.

This simple example shows that when non-equivalence holds, the selection of different priors can result in increasingly larger differences in DL when the system size grows. In other words, fine-tuning the prior can result in a significant gain - or, conversely, a loss - in terms of compression, particularly for large-scale systems. 

\section{Conclusion}

This study explored the implications of ensemble non-equivalence in the context of MDL-based model selection. We specifically compared the NML-based description lengths of canonical and microcanonical maximum-entropy models. The NML approach enables a comparison of model description lengths derived from the same underlying principle without requiring prior assumptions about the parameters. 

We found that, although microcanonical models fit the data better than their canonical counterparts, they are always more complex. Determining the best-scoring model - i.e., the one that minimizes the description length - is not straightforward, as it depends on the observed values of the sufficient statistics. Specifically, we found that the canonical model performs best when its fit to the observed data exceeds its average fit across all possible values of the sufficient statistics.

Additionally, we explicitly compared the description lengths of canonical and microcanonical models for dense $n\times m$ binary matrices in two cases of interest: one with a global constraint, such as the sum of the matrix entries, and another with $n$ local constraints, such as the sum of the entries in each row. In the former case, the model is homogeneous, where the probability of an entry being 1 is the same for all entries, while in the latter, the model becomes heterogeneous, with the probability of an entry being 1 depending on the specific row. In this second scenario, the canonical and microcanonical ensembles are no longer equivalent.

By comparing the description lengths of the canonical and microcanonical formulations of these models, we found that the difference remains finite when a global constraint is enforced but grows linearly with system size when local constraints are imposed. Our findings highlight the impact of choosing hard or soft constraints on data compression, leading us to conclude that canonical and microcanonical models should be treated as distinct from a model selection perspective in the presence of ensemble non-equivalence. Furthermore, they suggest a new definition of model equivalence based on MDL, which requires further investigation.

In addition, we investigated the relationship between NML-based and Bayesian description lengths. We defined NML-optimal priors as those that, when applied in a Bayesian framework, result in description lengths that asymptotically match the NML-based approach. While the Jeffreys prior is well known to be NML-optimal for canonical models under a single global constraint, we show that this optimality does not hold when local constraints are imposed. Indeed, in such cases, the difference between the corresponding description lengths increases with system size. This observation serves as a caution to practitioners, suggesting that the two approaches (NML and Jeffreys priors) should not be treated as interchangeable in the presence of ensemble non-equivalence. It also raises the question of whether an NML-optimal prior exists when local constraints are enforced.

Lastly, we showed that different priors can result in largely different description lengths when local constraints are involved. This finding underscores the importance for practitioners of carefully selecting priors, since the choice of prior could greatly impact the accuracy and effectiveness of model selection. 

\section*{Acknowledgements}
This work is supported by the European Union - NextGenerationEU - National Recovery and Resilience Plan (Piano Nazionale di Ripresa e Resilienza, PNRR), projects `SoBigData.it - Strengthening the Italian RI for Social Mining and Big Data Analytics' - Grant IR0000013 (n. 3264, 28/12/2021) (\url{https://pnrr.sobigdata.it/}); and ``Reconstruction, Resilience and Recovery of Socio-Economic Networks'' RECON-NET EP\_FAIR\_005 - PE0000013 ``FAIR'' - PNRR M4C2 Investment 1.3.

\appendix
 \section{Derivation of the microcanonical distribution from entropy maximization}
\label{micro_as_MEM}
It is well known that canonical maximum-entropy models can be derived by maximizing the Shannon entropy subject to soft constraints~\cite{jaynes}. Here, we show that microcanonical models can also be seen as the solution to an entropy maximization problem under hard constraints, as a direct consequence of the Shannon-Khinchin axioms that uniquely identify Shannon entropy~\cite{Khinchin57, somazzi23}.

Let $\mathcal{G}$ denote the set of all configurations, and let $\mathcal{G}_{\text{valid}} \subset \mathcal{G}$ denote the subset of configurations that exactly satisfy the hard constraints. Consider the problem of maximizing the Shannon entropy
\begin{equation}
S[P] = -\sum_{\mathbf{G} \in \mathcal{G}} P(\mathbf{G}) \log P(\mathbf{G})
\end{equation}
over probability distributions $P(\mathbf{G})$ supported on  $\mathcal{G}$, under the requirement that $ P(\mathbf{G}) = 0 $ for all $ \mathbf{G} \notin \mathcal{G}_{\text{valid}} $. That is, configurations violating the constraints are assigned zero probability.

The third Shannon-Khinchin axiom states that $S[P]$ is expansible, i.e., it does
not change if an outcome with zero probability is added. This implies that, in this case, maximizing entropy over the full space $\mathcal{G}$ with support restricted to $\mathcal{G}_{\text{valid}}$ is equivalent to maximizing entropy over $\mathcal{G}_{\text{valid}}$ only. Then, according to the second Shannon-Khinchin axiom, which states that entropy is maximized by the uniform distribution in the absence of further constraints, the solution is the uniform distribution over $\mathcal{G}_{\text{valid}}$:
\begin{equation}
P_{\text{mic}}(\bm{x}) = \begin{cases}
\frac{1}{\abs{\mathcal{G}_{\text{valid}}}} & \text{if }  \mathbf{G} \in \mathcal{G}_{\text{valid}}, \\
0 & \text{else}.
\end{cases}
\end{equation}
With this derivation, canonical and microcanonical ensembles emerge as distinct solutions to the same entropy maximization principle, differing only in how the constraints are imposed.

\section{Asymptotic expansions}

Here we derive the asymptotic expansions used to obtain the asymptotic results in section \ref{app}.

\subsection{One global constraint} \label{appendix_er}
We start by deriving an asymptotic expansion for the microcanonical log-likelihood:
\begin{align}
\ln \mathcal{L}_{\text{mic}}(l) &= - \ln \binom{nm}{l} \nonumber \\ &= - \ln (nm)! + \ln(nm-l)! +\ln l!.
\end{align}
In the dense regime, $l$ is of order $\Theta(n)$ and Stirling's approximation
\begin{equation}
\ln x! = x\ln x -x +\frac{1}{2} \ln (2\pi x) + \Theta(1/x)
\end{equation}
can be used to evaluate all the factorials above, yielding
\begin{multline}
\ln \mathcal{L}_{\text{mic}}(l) = - nm \ln (nm) + (nm-l)\ln(nm -l) \\+ l \ln l -\frac{1}{2}\ln \left(\frac{nm}{2\pi l (nm-l)}\right) + o(1), 
\end{multline}
which can be further expressed as a function of the density $p= l/nm$:  
\begin{multline}
    \ln \mathcal{L}_{\text{mic}}(p) = - nm \cdot h(p) + \frac{1}{2}\ln (nm) \\ + \frac{1}{2} \ln (2\pi p(1-p)) + o(1).
\end{multline}
By combining this expression with the canonical log-likelihood
\begin{equation}\label{as_can_l_er}
 \ln \mathcal{L}_{\text{can}}(p) = -nm\cdot h(p),
\end{equation}
we obtain the asymptotic log-likelihood difference of equation (\ref{as_delta_l_er}).

Similarly, we combine the asymptotic expansion of the microcanonical complexity
\begin{align}
    \text{COMP}_\text{mic} &= \ln(nm+1) =\ln (nm) + \ln \left(1 + \frac{1}{nm}\right) \nonumber \\ &= \ln (nm) + o(1),
\end{align}
which follows from $\ln\left(1+\frac{1}{x}\right) = \Theta(1/x)$, and the following asymptotic expansion of the canonical complexity
\begin{equation}\label{as_can_comp_er}
    \text{COMP}_\text{can} = \frac{1}{2} \ln (nm) + \frac{1}{2}\ln \left(\frac{\pi }{2}\right) +o(1).
\end{equation}
The latter is easily derived by asymptotically expanding the following approximation
\begin{multline}
\label{stan_approx}
\text{COMP}_\textrm{can}^\text{NML}\simeq\ln\left[\sqrt{\frac{\pi nm}{2}}+\frac{2}{3}+\frac{\sqrt{2\pi}}{24\sqrt{nm}}-\frac{4}{135nm}\right. \\+\left.\frac{\sqrt{2\pi}}{576\sqrt{(nm)^3}}+\frac{8}{2835(nm)^2}\right],
\end{multline}
provided in~\cite{staniczenko2014}, where the authors show it to be very precise already for small values of the total number of entries $nm$. 
Finally, equation (\ref{as_delta_c_er}) is obtained by comparing the two asymptotic complexities. Notice that the asymptotic formula (\ref{as_can_comp_er}) for the canonical complexity represents a well-known result, as it can be derived directly by applying the asymptotic formula (\ref{as_formula_nml}) to the case of i.i.d. Bernoulli random variables.  

\subsection{One-sided local constraints}

\label{appendix_pcm}
Here, we consider the regime in which $m =\Theta(n)$. Similarly to the previous case, we need to expand the microcanonical likelihood 
\begin{align}
    \ln\mathcal{L}_\textrm{mic}(\mathbf{r}) &= - \sum_{i=1}^n \ln \binom{m }{r_i} \nonumber \\ &= -n\ln m! + \sum_{i=1}^n \ln(m - r_i)! + \ln r_i!.
\end{align}
In the dense regime, all $r_i$'s are of order $\Theta(n)$, and we could again apply Stirling's formula to the factorials above. Nevertheless, because of the factor $n$ ahead of everything, Stirling's formula is not enough, and we turn to Stirling's series~\cite{stirlingseries}
\begin{equation}
\ln x! = x\ln x -x +\frac{1}{2} \ln (2\pi x) + \frac{1}{12x} + \Theta(1/x^3).
\end{equation}
The resulting asymptotic expression can be expressed as a function of the row densities $p_i = r_i/m$:
\begin{multline}
\ln\mathcal{L}_\textrm{mic}(\mathbf{r}) = -m \sum_{i=1}^n h(p_i)   +\frac{n}{2}\ln m  \\+ \frac{1}{2}\sum_{i=1}^n [\ln (2\pi  p_i(1-p_i))] -\frac{n}{12m} \\+ \frac{1}{12m}\sum_{i=1}^n\left[\frac{1}{p_i(1-p_i)}\right] +o(1).
\end{multline}
By combining this expression with the canonical log-likelihood
\begin{equation}
 \ln \mathcal{L}_{\text{can}}(\mathbf{p}) = -m\sum_{i=1}^n h(p_i),
\end{equation}
one obtains the asymptotic log-likelihood difference of equation (\ref{delta_logl_pcm}).

Similarly, the asymptotic expansion of the microcanonical complexity
\begin{align}
    \text{COMP}_\text{mic} &= n\ln(m+1) =\ln m + n\ln \left(1 + \frac{1}{m}\right)\nonumber \\  &= n\ln m + \frac{n}{m} + o(1),
\end{align}
which follows from $\ln\left(1+\frac{1}{x}\right) = \frac{1}{x} + \Theta(1/x^2)$, is compared to the following asymptotic expansion of the canonical complexity, based on (\ref{stan_approx})
\begin{multline}
\text{COMP}_\text{can} = \frac{n}{2}\ln m + \frac{n}{2}\ln\frac{\pi}{2} + \frac{2}{3}\sqrt{\frac{2}{\pi}}\cdot \frac{n}{\sqrt{m}} \\ + \left(\frac{1}{12}-\frac{4}{9\pi}\right) \cdot \frac{n}{m}+o(1)
\end{multline}
to express (\ref{comp_pcm}). 
The asymptotic expression (\ref{delta_comp_pcm}) is obtained as the difference between the two asymptotic complexities.

\section{NML and Bayes}

\subsection{Bayesian-Jeffreys description lengths}

\label{appendix_jeff}
In what follows, we compute the Bayesian description length $\text{DL}^{\text{B, Jeffreys}}_{\text{can}}$ of the canonical model obtained by constraining the sum $l$ over the matrix. As already stated, this model is equivalent to modeling $nm$ i.i.d. Bernoulli variables, and this result can be found in Example 8.3 of~\cite{grunwald_book}. In the paper, we extend this result to the case of $n$ one-sided local constraints. 

First, we compute Jeffreys prior. For a Bernoulli variable, the Fisher information for the parameter $p$ is $ \mathbf{I}(p)= p^{-1}(1-p)^{-1}$ and $\int_0^1 \sqrt{\mathbf{I}(p)} = \pi$. Thus, the Jeffreys prior reads
\begin{equation}
    J(p) = \frac{1}{\pi \sqrt{p(1-p)}}
\end{equation}
and the Bayesian-Jeffreys $\text{DL}$ is
\begin{multline}
\text{DL}^{\text{Bayes-Jeffreys}}_{\text{can}}(\mathbf{G}^*) = -\ln \int_0^1 P_{\text{can}}(\mathbf{G}^*; p)J(p)dp \\  = -\ln\frac{1}{\pi}\int_0^1 p^{l^* -\frac{1}{2}} (1-p)^{nm - l^* -\frac{1}{2}} dp.
\end{multline}
The integral above is the Beta function
\begin{equation}
   B(x, y) = \int_{0}^1 t^{x-1}(1-t)^{y-1}dt = \frac{\Gamma(x)\Gamma(y)}{\Gamma(x+y)},
\end{equation}
computed for $x = l^*+ \frac{1}{2}$ and $y=nm -l^* + \frac{1}{2}$, where $\Gamma(x)$ is the Gamma function
\begin{equation}
\Gamma(x) = \int_{0}^{+\infty} t^{x-1} e^{-t}dt.
\end{equation}
Thus
\begin{multline}
\text{DL}^{\text{Bayes-Jeffreys}}_{\text{can}}(\mathbf{G}^*) = \\ =  -\ln\frac{\Gamma\left(l^* + \frac{1}{2}\right)\Gamma\left(nm - l^* + \frac{1}{2}\right)}{\pi (nm)!},
\end{multline}
which corresponds to Equation (\ref{dl_jeff_er}). By employing the Stirling series, this expression can be expanded asymptotically as 
\begin{multline}\label{ap_as_jef_can_pcm}
\text{DL}^{\text{Bayes-Jeffreys}}_{\text{can}}(\mathbf{G}^*) = nm \cdot h\left(\frac{l^*}{nm}\right) \\ + \frac{1}{2}\ln \frac{\pi nm}{2} +\Theta\left(1/nm\right).
\end{multline}
The last expression is compared with the asymptotic expansion of the canonical NML description length $\text{DL}^\text{NML}_{\text{can}}$, obtained as the difference between the asymptotic expansion of (\ref{stan_approx}) and the canonical log-likelihood: 
\begin{multline}\label{ap_as_nml_can_pcm}
\text{DL}^\text{NML}_{\text{can}}(\mathbf{G}^*) = nm \cdot h\left(\frac{l^*}{nm}\right) \\ + \frac{1}{2}\ln \frac{\pi nm}{2} + \frac{2}{3} \sqrt{\frac{2}{\pi n m}} +o(1).
\end{multline}
Finally, Equation (\ref{as_diff_nml_jeff}) is obtained as the difference between (\ref{ap_as_nml_can_pcm}) and (\ref{ap_as_jef_can_pcm})

\subsection{Proving identities} \label{ap_identities}

In what follows, we prove identities (\ref{nml_can_bayesian_mic}) and (\ref{nml_mic_bayesian_can}) of section \ref{nml_bayes}. \\

\noindent\textit{Proving Identity (\ref{nml_can_bayesian_mic}).} We begin with identity (\ref{nml_can_bayesian_mic}), namely:

\begin{equation}
\text{DL}_{\text{can}}^{\text{NML}} = \text{DL}_{\text{mic}}^{\text{Bayes-Rissanen}},
\end{equation}
where the canonical prior $\hat{W}$ inducing the Bayes-Rissanen description length is expressed as
\begin{align}
\hat{W}(\mathbf{c}) &= \frac{\sum_{\mathbf{G}: \mathbf{c}(\mathbf{G}) = \mathbf{c}} P_{\text{can}}(\mathbf{G};\hat{\bm{\theta}}(\mathbf{G}) }{\sum_{\mathbf{G}}P_{\text{can}}(\mathbf{G};\hat{\bm{\theta}}(\mathbf{G}))}
\nonumber \\ &= \frac{\Omega(\mathbf{c})\mathcal{L}_\text{can}(\mathbf{c})}{\sum_{\mathbf{G}} \mathcal{L}_{\text{can}}(\mathbf{G})}. 
\end{align}
We recognize the canonical complexity in the denominator of $\hat{W}$. Thus, putting this expression in the Bayesian microcanonical description length, we get
\begin{align}
\text{DL}_{\text{mic}}^{\text{Bayes-Rissanen}}(\mathbf{G}^*)&=\ln \Omega(\mathbf{c}^*) - \ln \hat{W}(\mathbf{c}^*) \\
\nonumber &= - \ln \mathcal{L}_\text{can}(\mathbf{c}) + \text{COMP}_\text{can} \\
\nonumber &= \text{DL}_{\text{can}}^{\text{NML}}(\mathbf{G}^*),
\end{align}
which proves the identity.\\

\noindent\textit{Proving Identity (\ref{nml_mic_bayesian_can}).} We will now prove identity (\ref{nml_mic_bayesian_can}), namely:

\begin{equation}
\text{DL}_{\text{mic}}^{\text{NML}} = \text{DL}_{\text{can}}^{\text{Bayes-Uniform}},
\end{equation}
\noindent for the specific maximum-entropy models considered in this work, starting from the case of one global constraint. The uniform prior on the mean-value parameter $p$ reads
\begin{equation}
    w^{\text{Uniform}}(p) = 1 \quad \text{for } p\in[0,1].
\end{equation}
Putting this prior in the Bayesian description length yields:
\begin{align}
\text{DL}_{\text{can}}^{\text{Bayes-Uniform}}(\mathbf{G}^*) &= -\ln \int_0^1 P_\text{can}(\mathbf{G}^*; p) dp \nonumber \\
    \nonumber &= -\ln \int_0^1 p^{l^*} (1-p)^{nm - l^*} dp \\
    \nonumber &= \ln \binom{nm}{l^*} +\ln (nm+1)\\
    &=\text{DL}_{\text{mic}}^{\text{NML}}(\mathbf{G}^*),
\end{align}
where the integral is computed by integrating by parts $l^*$ times. This proves the identity for the case of one global constraint.

The identity holds as well for the case of $n$ one-sided local constraints, with the uniform prior reading
\begin{equation}
w^{\text{Uniform}}(\mathbf{p}) = 1 \quad \text{for } \mathbf{p}\in[0,1]^n.
\end{equation}
Indeed, we have that
\begin{align}
\nonumber \text{DL}_{\text{can}}^{\text{Bayes-Uniform}}(\mathbf{G}^*) &= -\ln \int_0^1 P_\text{can}(\mathbf{G}^*; p) dp \\
    \nonumber &= -\sum_{i=1}^n\ln  \int_0^1 p_i^{r_i^*} (1-p)^{m - r_i^*} dp_i \\
    \nonumber &= \sum_{i=1}^n\ln \binom{m}{r_i^*} +n\ln (m+1)
    \nonumber \\
&=\text{DL}_{\text{mic}}^{\text{NML}}(\mathbf{G}^*),
\end{align}
which proves the identity for the case of $n$ one-sided local constraints.

\bibliography{biblio}{}

\bibliographystyle{unsrt}

\end{document}